\documentclass[aps,prb,amsmath,amssymb,longbibliography,superscriptaddress,twocolumn]{revtex4-2}
\usepackage{braket}
\usepackage{fontenc}
\usepackage{graphicx}
\usepackage{hyperref}
\usepackage{bbm,color,ulem}
\DeclareMathAlphabet{\mathpzc}{OT1}{pzc}{m}{it} 
\begin{document}
\title{A generalized model of the noise spectrum of a two-level fluctuator in the presence of an electron subbath}
\author{Robert E.\ Throckmorton}
\author{S.\ \surname{Das Sarma}}
\affiliation{Condensed Matter Theory Center and Joint Quantum Institute, Department of Physics, University of Maryland, College Park, Maryland 20742-4111 USA}
\date{\today}
\begin{abstract}
The work of Ahn {\it et.~al.} [Phys.~Rev.~B {\bf 103}, L041304 (2021)] derives the noise power spectrum of a two-level fluctuator (TLF) in the case where it interacts only with a subregion of a full electron bath and thus is subject to a fluctuating temperature.  However, Eq.~(1), which gives the variance of the subbath temperature in terms of the heat capacity, in that work carries the implicit assumption that the heat capacity of this subbath may be taken to be a constant, which is a good approximation at higher temperatures, but breaks down at lower temperatures.  We thus extend this work to the case in which the fact that the electronic heat capacity of a two-dimensional electron gas (2DEG) $C_V\propto T_0$, where $T_0$ is the temperature of the full 2DEG, rather than constant in temperature, is fully taken into account.  We show that, at low temperatures, the resulting power spectrum of the noise $S(\omega,T_0)\propto e^{-C/T_0^{3/8}}$, in contrast to $S(\omega,T_0)\propto e^{-C'/T_0^{1/3}}$ as found previously, where $C$ and $C'$ are constants.  We also compare the numerical results that one would obtain from the two models and find that our results for $S(\omega,T_0)$ can differ from those of Ahn by several orders of magnitude at low temperatures.  Finally, we perform a fit of the spectra of two TLFs to experimental data from Connors {\it et.~al.} [Phys.~Rev.~B {\bf 100}, 165305 (2019)] using our results and find excellent agreement with the data over most of the temperature range considered in the experiment.
\end{abstract}
\maketitle

\section{Introduction}
One of the leading platforms for the realization of qubits, and thus an eventual quantum computer, is the semiconductor-based electron spin qubit.  There are multiple types of semiconductor-based spin qubits, though all of them are composed of one or more electrons, each trapped in a quantum dot.  The major advantages of these semiconductor-based electron spin qubits over the other leading platforms, superconducting and ion trap qubits, are their smaller size (on the order of nanometers) and fast electrical control.  However, the main disadvantage that this platform faces compared to the others is the comparatively low fidelity of gate operations.  Nevertheless, these systems have been of great interest experimentally due to their potential advantages, with demonstrations of single-qubit \cite{s41928-019-0234-1,acs.nanolett.0c02397,acs.nanolett.0c04771} and two-qubit\cite{nnano.2014.216,nature15263,s41586-019-1197-0,science.aao5965,nature25766,PhysRevX.9.021011,sciadv.abn5130,s41586-021-04182-y,s41586-021-04273-w} gates.  Some of these experiments in particular have demonstrated two-qubit gate fidelities at or above $99\%$\cite{sciadv.abn5130,s41586-021-04182-y,s41586-021-04273-w}, putting error correction techniques within reach in semiconductor-based electron spin qubits.

One of the main challenges facing semiconductor-based electron spin qubits is magnetic and electronic noise in the qubit systems, which is responsible for the lower gate fidelities in such qubits compared to other platforms.  Therefore, investigations of noise in these systems and the development of methods to mitigate its effects are critical to the eventual development of a working quantum computer built from semiconductor-based electron spin qubits.  Both theoretical and experimental investigations of noise in semiconductor systems have in fact been undertaken, some outside of the context of semiconductor-based electron spin qubits.  A work by Dutta {\it et. al.} \cite{PhysRevLett.43.646} and a later work by Dutta and Horn \cite{RevModPhys.53.497} showed that, given an approximately uniform distribution of activation energies, the total noise power spectrum produced by an ensemble of two-level fluctuators (TLFs) follows a $1/f$ distribution.  Experiments \cite{PhysRevLett.121.076801} later found a $1/f$ distribution in an experimental Si-based spin qubit system at low frequencies, but further experiments \cite{TNS.2015.2405852,PhysRevApplied.10.044017,PhysRevB.100.165305,NC.13.940} found deviations from $1/f$ behavior in the overall spectrum, suggesting an ensemble of TLFs with a nonuniform distribution of activation energies.  On the theoretical side, one work \cite{aa7648} describes the use of a qubit to measure the noise power spectrum.  In addition to these investigations of the noise power spectrum itself, a number of theoretical works have characterized the effects of noise on information loss in a qubit \cite{PhysRevB.105.245413} and considered how to mitigate the effects of noise on single- and two-qubit gates \cite{PhysRevA.90.042307,PhysRevB.93.121407,PhysRevB.94.045435,PhysRevB.95.085405,npjQI.4.62,PhysRevB.99.081301,PhysRevB.108.045305}.

Our work is focused in particular on extending the analysis of Ahn {\it et. al.}~\cite{PhysRevB.103.L041304}.  It had long been assumed that the observed $1/f$-like charge noise spectrum in semiconducting systems required an ensemble of two-level fluctuators (TLFs), which was the case considered in the work of Dutta {\it et. al.} \cite{PhysRevLett.43.646}.  However, Ahn {\it et. al.} showed that experimental data \cite{PhysRevB.100.165305} can be fit by just one or two TLFs if it is assumed that these TLFs interact only with small subregions, or subbaths, of the full electron bath in the semiconductor.  Even if the full two-dimensional electron gas (2DEG) is in thermal equilibrium, it is still possible for local fluctuations in energy, and thus temperature, to occur.  As a result, the temperatures of these subbaths will fluctuate, thus producing an effect similar to averaging over an ensemble of TLFs with different activation energies.

We note, however, that Eq.~(1) in this work, which gives the variance in the subbath temperature $\sigma_{sb}^2$ in terms of the heat capacity of the subbath $C_V$,
\begin{equation}
\sigma_{sb}^2=\frac{k_BT_0^2}{C_V}, \label{Eq:SigmaSB_CVConst}
\end{equation}
where $T_0$ is the temperature of the full electron bath, corresponding to the average energy $E_0$, carries the implicit assumption that the heat capacity of the subbath may be treated as a constant in temperature; in reality, the heat capacity is linear in temperature.  This equation can be derived from the standard relation between the variance in energy and the electronic heat capacity, $\sigma_E^2=k_BT_0^2C_V$.  If we can assume that $C_V$ is approximately a constant in temperature, then $\sigma_E=C_V\sigma_{sb}$, and Eq.~\eqref{Eq:SigmaSB_CVConst} immediately follows.  We expect this assumption to hold well for high temperatures (i.e., $T_0\gg\sigma_{sb}$), but it will break down at lower temperatures.  If the assumption that $C_V$ is constant in temperature breaks down, then $\sigma_E$ will no longer be proportional to $\sigma_{sb}$, and thus Eq.~\eqref{Eq:SigmaSB_CVConst} will no longer hold.  We thus perform an analysis that takes this variation in heat capacity with temperature into full account, finding significant differences from Ahn {\it et. al.}'s results for $S(\omega,T_0)$ for a single TLF at low temperatures.

We start from the assumption that the energy of a subbath has a Gaussian distribution, and we derive the corresponding temperature distribution.  From this, we derive the Gaussian approximation to this temperature distribution and determine the standard deviation of the subbath temperature $\sigma_{sb}$.  We show that, at low temperatures, $\sigma_{sb}\propto T_0^{3/4}$.  We then investigate the noise power spectrum $S(\omega,T_0)$.  We show, again at low temperatures, that $S(\omega,T_0)\propto e^{-C/T_0^{3/8}}$.  In contrast, Ahn {\it et. al.} finds that $\sigma_{sb}\propto\sqrt{T_0}$ and $S(\omega,T_0)\propto e^{-C'/T_0^{1/3}}$, respectively, in the same limit \cite{PhysRevB.103.L041304}.  Finally, we numerically calculate the noise power spectrum using both Ahn {\it et. al.}'s approach and our own approach.  We find that the results from each can differ by several orders of magnitude at low temperatures, but the overall qualitative picture remains the same---we still find $1/\omega^{0.9}$ behavior at low frequencies and $1/\omega^2$ at high frequencies, with a transitional region over which $S(\omega,T_0)$ remains flat.

We then compare our model to experimental data.  We perform a fit of two TLFs of different strengths, activation energies, and switching times to the data of Ref.~\cite{PhysRevB.100.165305}, which presents measurements of the noise spectra in two quantum dots.  We see that the fit to the data is excellent over a large temperature range, helping to illustrate that even just two TLFs can easily explain the observed noise spectrum in a given qubit.  We also note that the same activation energies and switching times produce such excellent fits to the spectra for both dots---only the strengths of the TLFs differ.  This implies that the same two TLFs are causing the noise in both dots, as the different strengths can be explained by the positions of the TLFs relative to the two dots.

The rest of the paper is organized as follows.  We derive the temperature distribution and determine the value of $\sigma_{sb}$ in a Gaussian approximation in Sec.~\ref{Sec:TempDist}.  We derive analytical results for the noise power spectrum at low temperature in Sec.~\ref{Sec:NoisePwrSpec}.  We then provide numerical results for it in Sec.~\ref{Sec:Comparison}, along with our comparison to the experimental data of Ref.~\cite{PhysRevB.100.165305}.  We finally give our conclusions in Sec.~\ref{Sec:Conclusion}.

\section{Temperature distribution} \label{Sec:TempDist}
We start with the assumption that the distribution of energies of the subbath may be taken to be a truncated Gaussian:
\begin{equation}
f_E(E)\propto e^{-(E-E_0)^2/2\sigma_E^2}, E\in [0,\infty).
\end{equation}
By adopting this distribution, we assume that the lowest energy of the subbath is zero.  For a 2DEG with (effective) electronic mass $m^\ast$, the heat capacity of the subbath as a function of temperature is
\begin{equation}
C_V=\alpha T, \label{Eq:2DEG_CV}
\end{equation}
where
\begin{equation}
\alpha=\frac{\pi m^\ast k_B^2A}{3\hbar^2}
\end{equation}
and $A$ is the area of the subbath that the TLF interacts with.  We can find the total energy of the subbath, obtaining
\begin{equation}
E(T)=\tfrac{1}{2}\alpha T^2. \label{Eq:EFuncOfT}
\end{equation}
With this, we can now derive the subbath temperature distribution $f_T(T)$ corresponding to the energy distribution $f_E(E)$:
\begin{equation}
f_T(T)\propto Te^{-(\alpha T^2-2E_0)^2/8\sigma_E^2}, T\in [0,\infty).
\end{equation}
We now introduce a temperature scale $T_0$, the temperature of the full electron bath, related to $E_0$ via Eq.~\eqref{Eq:EFuncOfT}:
\begin{equation}
E_0=\tfrac{1}{2}\alpha T_0^2.
\end{equation}
In terms of this temperature scale, $f_T(T)$ becomes
\begin{equation}
f_T(T)\propto Te^{-\alpha^2(T^2-T_0^2)^2/8\sigma_E^2}, T\in [0,\infty). \label{Eq:TempDistExact}
\end{equation}

We now want to find the relation between the effective standard deviation $\sigma_{sb}$ of this distribution around its maximum and that of the energy distribution, $\sigma_E$.  To this end, we approximate the temperature distribution as a Gaussian.  We begin by collecting the entire dependence of $f_T(T)$ on temperature in the exponent:
\begin{equation}
f_T(T)=Ce^{-g(T)},
\end{equation}
where
\begin{equation}
g(T)=\frac{\alpha^2(T^2-T_0^2)^2}{8\sigma_E^2}-\ln\left (\frac{T}{T_s}\right )
\end{equation}
and $T_s$ is an arbitrary temperature scale.  We first find the temperature $T^\ast$ at which this exponent is maximized.  If we let
\begin{equation}
x^\ast=\frac{\alpha (T^\ast)^2}{2\sigma_E}\text{ and }x_0=\frac{\alpha T_0^2}{2\sigma_E},
\end{equation}
then the equation giving $T^\ast$ is
\begin{equation}
2x^\ast(x^\ast-x_0)=1.
\end{equation}
Since $x^\ast>0$, the sole physical solution is
\begin{equation}
x^\ast=\frac{x_0+\sqrt{x_0^2+2}}{2}.
\end{equation}
We can now obtain the temperature variance $\sigma_{sb}^2$ from the second derivative of $g(T)$:
\begin{equation}
\frac{1}{\sigma_{sb}^2}=g''(T=T^\ast)=\frac{2\alpha}{\sigma_E}\sqrt{x_0^2+2}, \label{Eq:SBVar_SecDer}
\end{equation}
or
\begin{equation}
\sigma_{sb}^2=\frac{\sigma_E}{2\alpha\sqrt{x_0^2+2}}=\frac{\sigma_E^2}{\alpha\sqrt{\alpha^2 T_0^4+8\sigma_E^2}}. \label{Eq:SBVar}
\end{equation}
The relation, Eq.~\eqref{Eq:SBVar_SecDer}, simply follows from the Taylor expansion of $g(T)$ around $T=T^\ast$:
\begin{equation}
g(T)=g(T^\ast)+\tfrac{1}{2}g''(T^\ast)(T-T^\ast)^2+\cdots.
\end{equation}
If we truncate this expansion at second order, we obtain a Gaussian distribution with a variance given by Eq.~\eqref{Eq:SBVar}.

We now consider two extreme cases.  If $T_0\gg\sqrt{2\sigma_E/\alpha}$ (i.e., the high-temperature limit), then
\begin{equation}
\sigma_{sb}\approx\frac{\sigma_E}{\alpha T_0}=\frac{\sigma_E}{C_V}. \label{Eq:STHighTemp}
\end{equation}
On the other hand, if $T_0\ll\sqrt{2\sigma_E/\alpha}$ (i.e., the low-temperature limit), then
\begin{equation}
\sigma_{sb}^2\approx\frac{\sigma_E}{2\sqrt{2}\alpha}. \label{Eq:STLowTemp}
\end{equation}

We can relate $\sigma_E$ to the temperature $T_0$ as follows.  The heat capacity is related to the variance in energy $\sigma_E^2$ by
\begin{equation}
C_V=\frac{\sigma_E^2}{k_BT_0^2}. \label{Eq:CVSERel}
\end{equation}
Using this relation, we obtain
\begin{equation}
\sigma_E^2=k_B\alpha T_0^3,
\end{equation}
or
\begin{equation}
\sigma_E=\sqrt{k_B\alpha}T_0^{3/2}.
\end{equation}
Using the low-temperature approximation, Eq.~\eqref{Eq:STLowTemp}, we obtain
\begin{equation}
\sigma_{sb}=\left (\frac{k_B}{8\alpha}\right )^{1/4}T_0^{3/4}.
\end{equation}

To contrast with Ahn {\it et. al.}'s results, let us briefly consider the case in which the specific heat is taken to be a constant, so that $E=C_VT$.  In this case, we would find that $f_T(T)$ is exactly Gaussian, with a standard deviation of $\sigma_{sb}=\sigma_E/C_V$.  We would then find from Eq.~\eqref{Eq:CVSERel} that
\begin{equation}
\sigma_{sb}^2=\frac{k_BT_0^2}{C_V},
\end{equation}
which is just Eq.~(1) in Ref.~\cite{PhysRevB.103.L041304}.  If we were to substitute the formula for $C_V$, Eq.~\eqref{Eq:2DEG_CV}, into this equation, we would find that $\sigma_{sb}\propto\sqrt{T_0}$.  This would, in fact, correspond to the high-temperature limit, Eq.~\eqref{Eq:STHighTemp}, thus showing that the implicit approximation made in Ref.~\cite{PhysRevB.103.L041304} works well at higher temperatures, as expected.

\section{Noise power spectrum} \label{Sec:NoisePwrSpec}
We now turn our attention to the noise power spectrum $S(\omega,T_0)$.  The power spectrum for a single TLF is
\begin{equation}
S_\text{TLF}(\omega,T)=\frac{4\Delta^2\tau}{1+\omega^2\tau^2},
\end{equation}
where $\Delta$ is the strength of the fluctuator and $\tau$ is the characteristic switching time.  We assume that the switching time has an ``activated'' dependence on temperature, i.e., $\tau=\tau_0 e^{\epsilon/k_BT}$, with $\epsilon$ being the energy difference between the two states of the fluctuator and $T$ being the temperature of the subbath in the vicinity of the fluctuator.

\begin{figure*}
	\centering
	\includegraphics[width=0.33\linewidth]{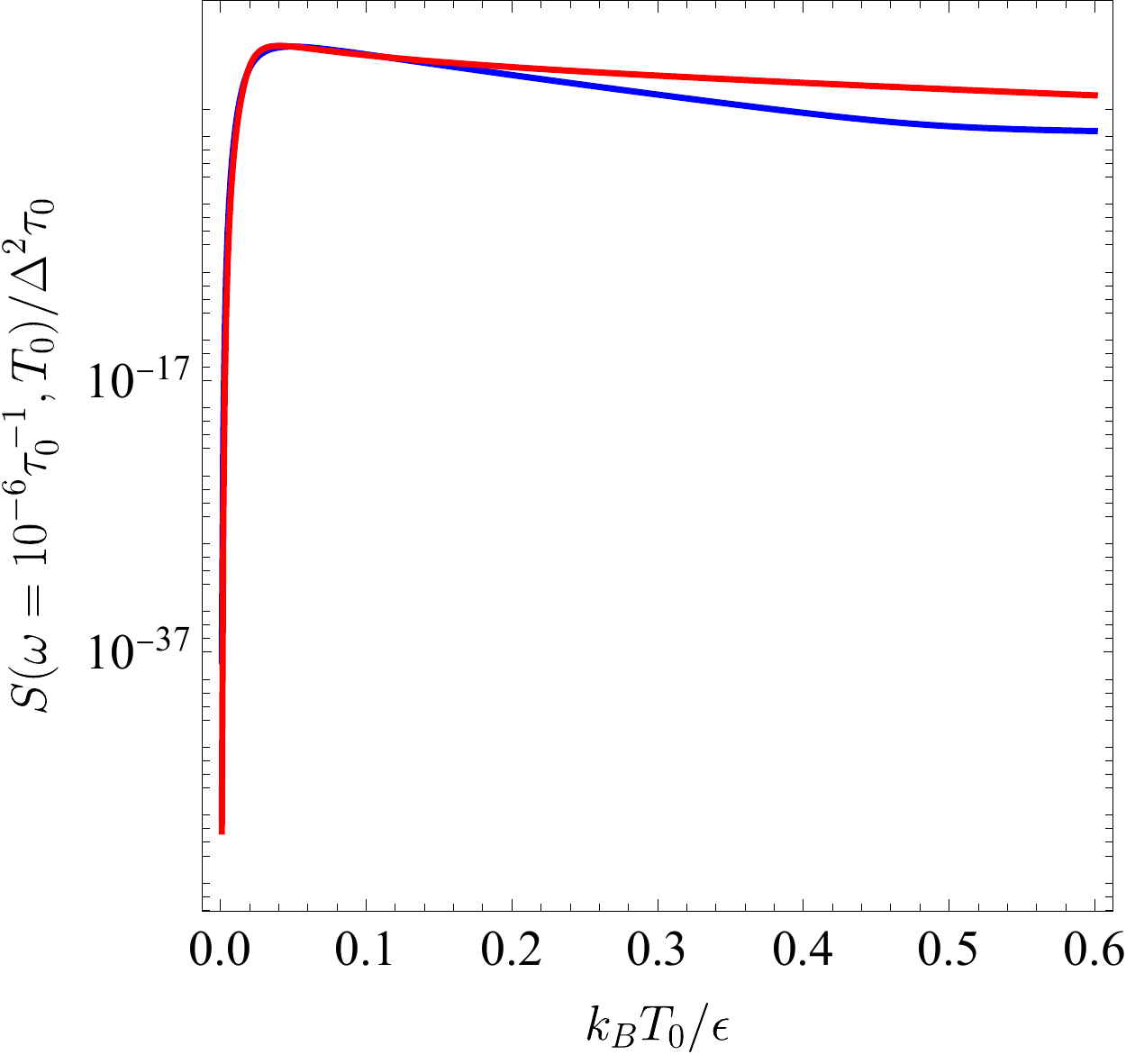}
	\includegraphics[width=0.33\linewidth]{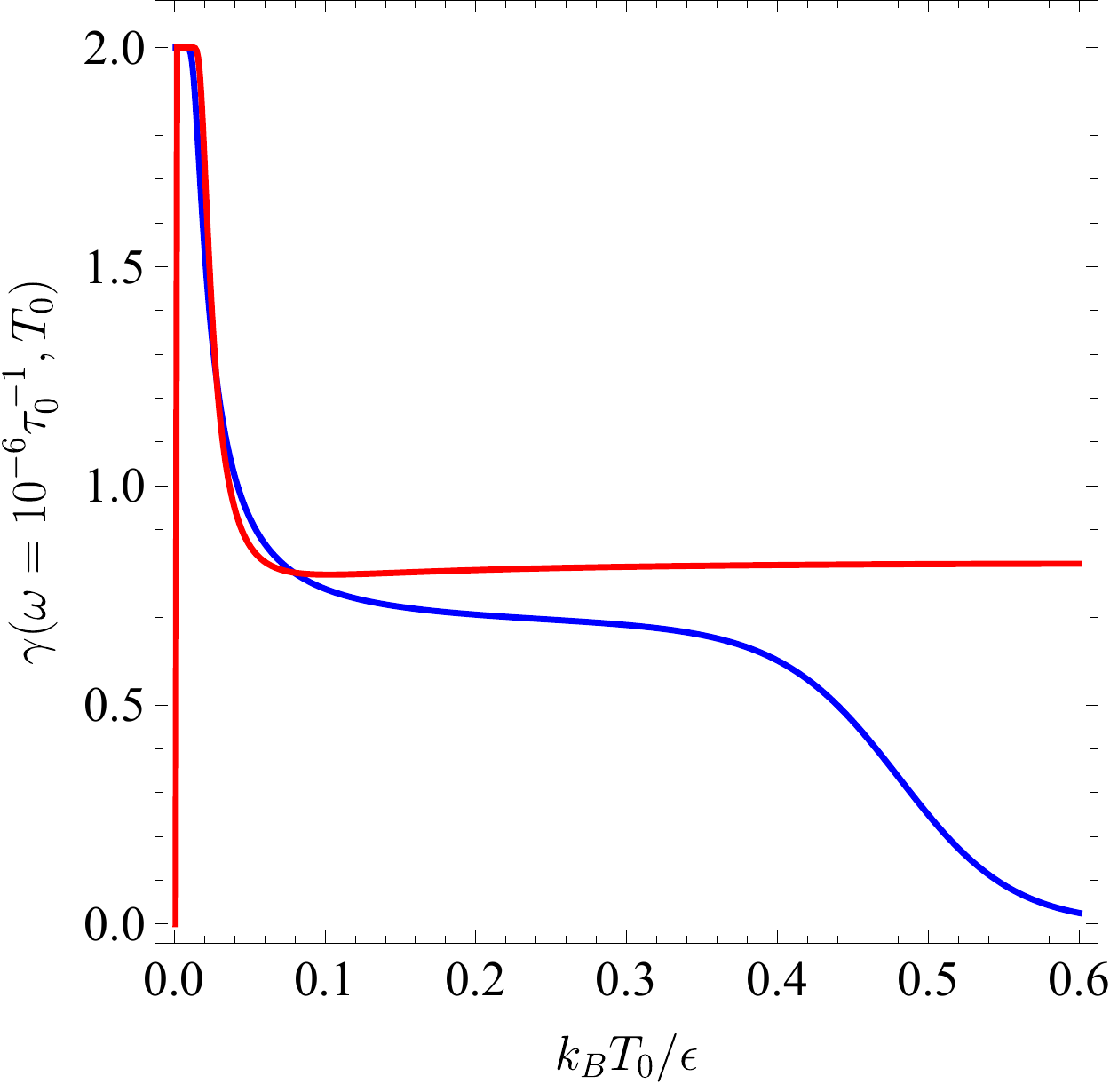}
	\caption{Plot of the noise power spectrum $S(\omega,T_0)$ (left panel) and of $\gamma(\omega,T_0)$ (right panel) for both the exact temperature distribution, Eq.~\eqref{Eq:TempDistExact} (red curves) and the approximate distribution, Eq.~\eqref{Eq:TempDistGaussianApprox} (blue curves) as functions of the full electron bath temperature $T_0$ for constant frequency $\omega\tau_0=10^{-6}$.}
	\label{fig:ConstOmegam6_RT_ExactVsApprox}
\end{figure*}

We now consider the effect of temperature fluctuations in the subbath.  We will be working in the low-temperature limit, and thus we will approximate the temperature distribution using the effective Gaussian form derived above, i.e.,
\begin{equation}
f_T(T)=\sqrt{\frac{2}{\pi}}\frac{1}{1+\mbox{erf}(T^\ast/\sigma_{sb}\sqrt{2})}e^{-(T-T^\ast)^2/2\sigma_{sb}^2}. \label{Eq:TempDistGaussianApprox}
\end{equation}
The full noise power spectrum $S(\omega,T_0)$ is then given by
\begin{eqnarray}
S(\omega,T_0)&=&\int_0^\infty f_T(T)S_\text{TLF}(\omega,T)\,dT \cr
&=&\sqrt{\frac{2}{\pi}}\frac{4\Delta^2\tau_0}{1+\mbox{erf}(T^\ast/\sigma_{sb}\sqrt{2})}\int_0^\infty \frac{e^{-(T-T^\ast)^2/2\sigma_{sb}^2}e^{\epsilon/k_BT}}{1+\omega^2\tau_0^2 e^{2\epsilon/k_BT}}\,dT. \nonumber \\
\end{eqnarray}
To justify our Gaussian approximation, we now turn our attention to comparing the numerical results for $S(\omega,T_0)$ that we would obtain from our approximate Gaussian distribution, Eq.~\eqref{Eq:TempDistGaussianApprox} to those found from the exact distribution, Eq.~\eqref{Eq:TempDistExact}, as well as for $\gamma(\omega,T_0)$, which is given by 
\begin{equation}
	\gamma(\omega,T_0)=-\frac{\partial\ln{S}}{\partial\ln{\omega}}. \label{eq:GammaDef}
\end{equation}
We present a plot comparing the results from each distribution in Fig.~\ref{fig:ConstOmegam6_RT_ExactVsApprox}.  We see that the results agree well for low temperatures, thus showing that the approximation is indeed valid for sufficiently low temperatures (i.e., $k_BT/\epsilon<0.1$).  We also see that $\gamma(\omega,T_0)\approx 2$ at low temperatures for both the exact model and the Gaussian approximation, consistent with the approximate dependence that we will find shortly in Eq.~\eqref{Eq:SApprox}.  We expect the Gaussian approximation to work best at low temperatures because the temperature distribution $f_T(T)$ is narrower at lower temperatures than at higher temperatures ($\sigma_E\propto T_0^{3/2}$ and $\sigma_{sb}\propto T_0^{3/4}$), and thus varies much more quickly than $S_\text{TLF}(\omega,T)$ in the vicinity of $T^\ast$.

At low temperature, the exponentials $e^{\epsilon/k_BT}$ become very large, and thus we may make the approximation,
\begin{eqnarray}
S(\omega,T_0)&\approx& \sqrt{\frac{2}{\pi}}\frac{4\Delta^2}{\omega^2\tau_0}\frac{1}{1+\mbox{erf}(T^\ast/\sigma_{sb}\sqrt{2})} \cr
&\times&\int_0^\infty \exp\left [-\frac{(T-T^\ast)^2}{2\sigma_{sb}^2}-\frac{\epsilon}{k_BT}\right ]\,dT. \label{eq:SApprox}
\end{eqnarray}
To evaluate the integral, we approximate the integrand as a Gaussian itself:
\begin{equation}
\exp\left [-\frac{(T-T^\ast)^2}{2\sigma_{sb}^2}-\frac{\epsilon}{k_BT}\right ]\approx Be^{-(T-T^{\ast\ast})^2/2\sigma_\ast^2} \label{eq:SIntApprox}
\end{equation}
Following a similar procedure as before, we find that the integrand is maximized at a temperature of $T=T^{\ast\ast}$ given by
\begin{equation}
(T^{\ast\ast})^2(T^{\ast\ast}-T^\ast)=\frac{\sigma_{sb}^2\epsilon}{k_B}.
\end{equation}
If we work in the low-temperature limit
\begin{equation}
T^\ast\ll\left (\frac{\sigma_{sb}^2\epsilon}{k_B}\right )^{1/3}, \label{Eq:LowTemp_TStar}
\end{equation}
then we find that
\begin{equation}
T^{\ast\ast}\approx\tfrac{1}{3}T^\ast+\left (\frac{\epsilon}{\sqrt{8k_B\alpha}}\right )^{1/3}\sqrt{T^\ast}
\end{equation}
and
\begin{equation}
\frac{1}{\sigma_\ast^2}=\frac{1}{\sigma_{sb}^2}+\frac{2\epsilon}{k_B(T^\ast)^3}.
\end{equation}
We now recall our earlier result that, if $x^\ast=\frac{\alpha(T^\ast)^2}{2\sigma_E}$, then
\begin{equation}
x^\ast=\frac{x_0+\sqrt{x_0^2+2}}{2}.
\end{equation}
If we assume that $T_0\ll\frac{k_B}{\alpha}$, then this yields
\begin{equation}
T^\ast\approx\left (\frac{k_B}{2\alpha}\right )^{1/4}T_0^{3/4}.
\end{equation}
We thus find that the assumed low-temperature limit, defined by Eq.~\eqref{Eq:LowTemp_TStar}, is valid, since $T^\ast$ goes as $T_0^{3/4}$, while $(\sigma_{sb}^2\epsilon/k_B)^{1/3}$ goes as $\sqrt{T_0}$.

This, in turn, gives us
\begin{equation}
\sigma_\ast\approx\sqrt{\frac{k_B}{2\epsilon}}\left (\frac{k_B}{2\alpha}\right )^{3/8}T_0^{9/8}.
\end{equation}

We can now find the low-temperature behavior of the noise power spectrum $S(\omega,T_0)$.  Evaluating Eq.~\eqref{eq:SApprox} using the approximation in Eq.~\eqref{eq:SIntApprox}, we find that
\begin{equation}
S(\omega,T_0)=\frac{4\Delta^2}{\omega^2\tau_0}\frac{1}{1+\mbox{erf}(T^\ast/\sigma_{sb}\sqrt{2})}\frac{1}{\pi\sigma_\ast}B, \label{Eq:SApprox}
\end{equation}
where
\begin{equation}
B=\exp\left [-\frac{(T^{\ast\ast}-T^\ast)^2}{2\sigma_{sb}^2}-\frac{\epsilon}{k_BT^{\ast\ast}}\right ]\approx\frac{1}{e^2}\exp\left (-\frac{C}{T_0^{3/8}}\right )
\end{equation}
and
\begin{equation}
C=2^{5/8}\frac{\epsilon}{k_B}\left (\frac{k_B\alpha^7}{\epsilon^8}\right )^{1/24}.
\end{equation}
This dependence differs somewhat from that found in Ref.~\cite{PhysRevB.103.L041304}, $S(\omega,T_0)\propto e^{-C'/T_0^{1/3}}$.  We also note from Eq.~\eqref{Eq:SApprox} that $S(\omega,T_0)\propto 1/\omega^2$ at the low-temperature limit, consistent with our earlier numerical results for $\gamma$ (Fig.~\ref{fig:ConstOmegam6_RT_ExactVsApprox}).

Let us now turn to determining a formula for $\gamma$, defined in Eq.~\eqref{eq:GammaDef}, that generalizes Eqs.~(18) and (19) in Ref.~\cite{PhysRevB.103.L041304}.  Starting from the Dutta-Horn model,
\begin{equation}
S_\text{TLF}(\omega,T)=\frac{2\pi k_BT\Delta^2}{\omega}F(E_\omega),
\end{equation}
where $F$ is a function determined by the energy distribution of the two-level fluctuators and
\begin{equation}
E_\omega=-k_BT\ln(\omega\tau_0),
\end{equation}
we want to find the full power spectrum
\begin{equation}
S(\omega,T_0)=\frac{2\pi k_B\Delta^2}{\omega}\int_0^{\infty} TF(E_\omega)f_T(T)\,dT.
\end{equation}
Following the same basic procedure as in Ref.~\cite{PhysRevB.103.L041304} [except we expand $TF(E_\omega)$ around $T^\ast$], we obtain
\begin{widetext}
\begin{equation}
\gamma=1-\frac{1}{\ln(\omega\tau_0)}\left (\frac{\partial\ln{S}}{\partial\ln{T^\ast}}-1\right )\left\{1+\frac{\left [\left (\frac{\sigma_{sb}}{T^\ast}\right )^2-\frac{\sigma_{sb}}{T^\ast}\frac{d\sigma_{sb}}{dT^\ast}\right ](2F'(E_\omega^\ast)+E_\omega^\ast F''(E_\omega^\ast))}{\left [1-\left (\frac{\sigma_{sb}}{T^\ast}\right )^2+2\frac{\sigma_{sb}}{T^\ast}\frac{d\sigma_{sb}}{dT^\ast}\right ]F'(E_\omega^\ast)+\left [\left (\frac{\sigma_{sb}}{T^\ast}\right )^2+\frac{\sigma_{sb}}{T^\ast}\frac{d\sigma_{sb}}{dT^\ast}\right ]E_\omega^\ast F''(E_\omega^\ast)+\tfrac{1}{2}\left (\frac{\sigma_{sb}}{T^\ast}\right )^2(E_\omega^\ast)^2F'''(E_\omega^\ast)}\right\},
\end{equation}
\end{widetext}
where $E_\omega^\ast=E_\omega(T=T^\ast)$.  In the limit that the second- and higher-order derivatives of $F$ are negligible, we obtain
\begin{equation}
\gamma\approx 1-\frac{1}{\ln(\omega\tau_0)}\left (\frac{\partial\ln{S}}{\partial\ln{T^\ast}}-1\right )\frac{1+\left (\frac{\sigma_{sb}}{T^\ast}\right )^2}{1-\left (\frac{\sigma_{sb}}{T^\ast}\right )^2+2\frac{\sigma_{sb}}{T^\ast}\frac{d\sigma_{sb}}{dT^\ast}}.
\end{equation}
This is similar to Eq.~(19) in Ref.~\cite{PhysRevB.103.L041304}, except for the extra term, $2\frac{\sigma_{sb}}{T^\ast}\frac{d\sigma_{sb}}{dT^\ast}$, in the denominator.

\section{Comparison with the previous work} \label{Sec:Comparison}
We now determine how large an effect the considerations raised so far have on Ahn {\it el. al.}'s results \cite{PhysRevB.103.L041304}.  To this end, we numerically calculate $S(\omega,T_0)$ and $\gamma(\omega,T_0)$ for a single TLF using Ahn {\it el. al.}'s formulas and compare the results to those obtained from ours.  Throughout this calculation, we assume the following parameters, which are largely based on the fitting parameters found in Table I of Ref.~\cite{PhysRevB.103.L041304}: $\epsilon=0.216\,\text{meV}$, $\tau_0=8.397\,\text{ms}$, and $r=182.8\,\text{nm}$, where the area of the subbath is $A=\pi r^2$.  We also take $m^\ast=0.19m_0$, which is the transverse effective mass of electrons in silicon.  For the results obtained with our formulas, we use the exact temperature distribution, Eq.~\eqref{Eq:TempDistExact}, rather than the approximate Gaussian distribution; this is because the temperature range that we are considering includes temperatures outside the range of validity of the approximate distribution.

\begin{figure*}
	\centering
	\includegraphics[width=0.32\textwidth]{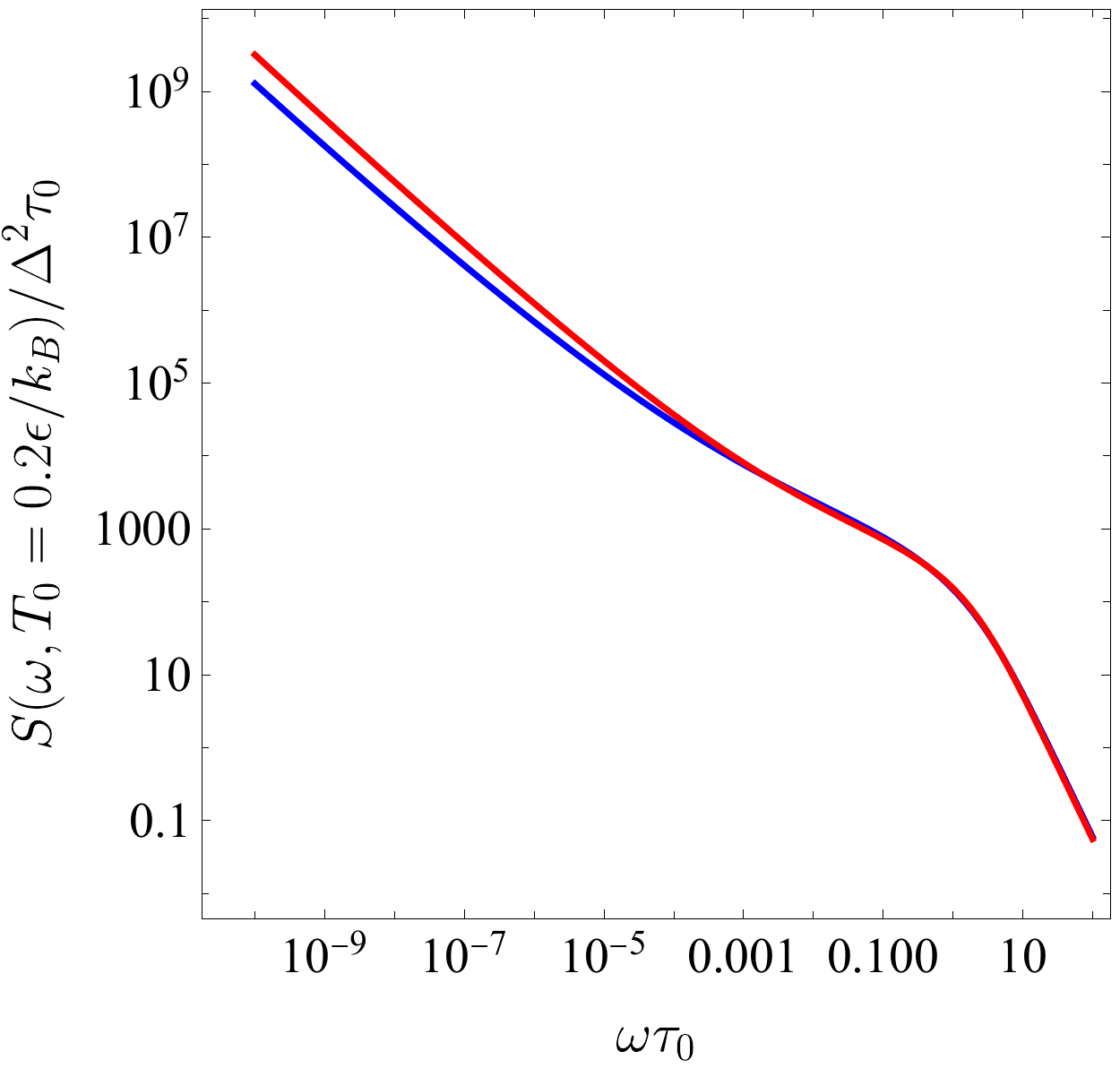}
	\includegraphics[width=0.32\textwidth]{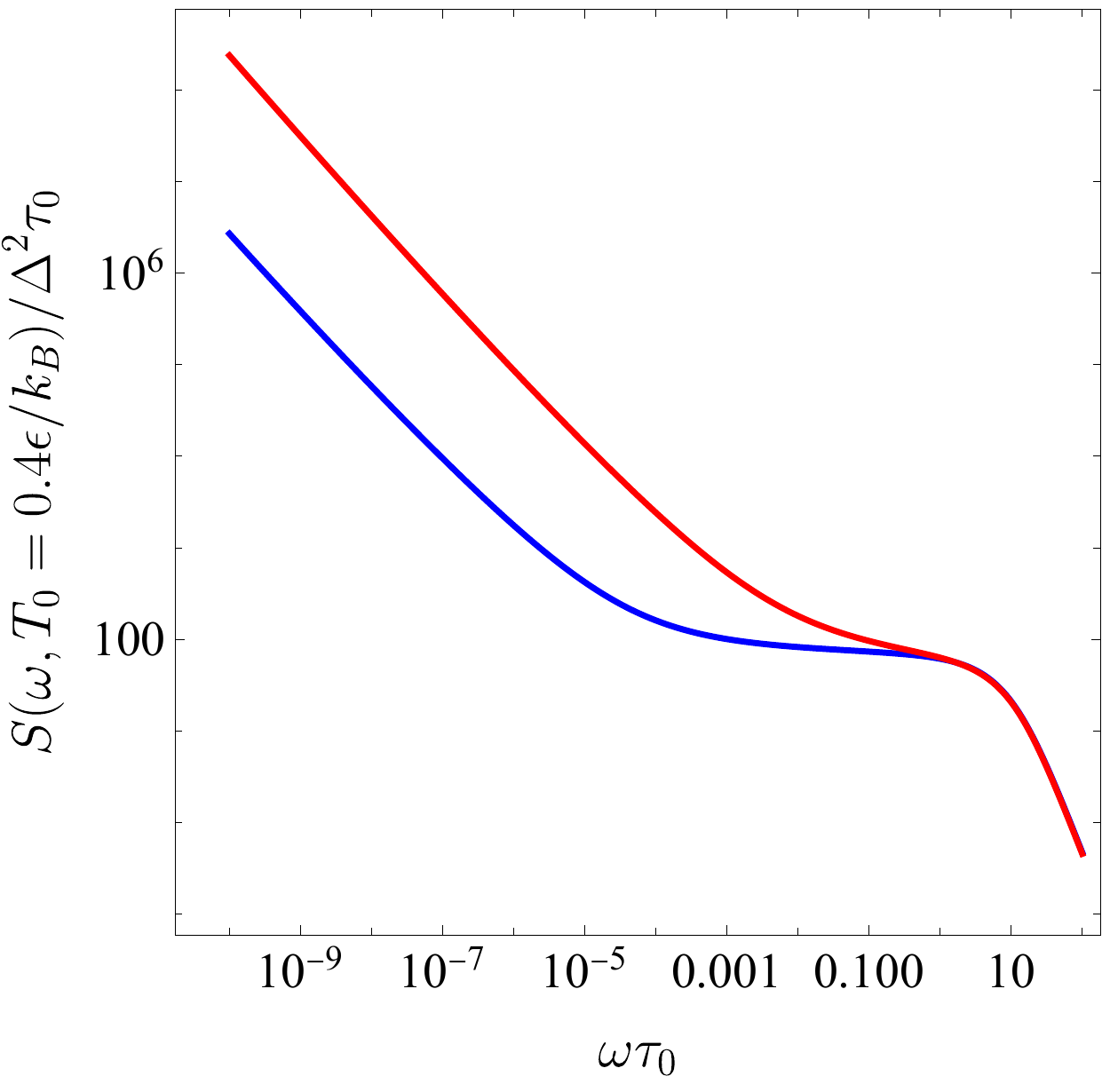}
	\includegraphics[width=0.32\textwidth]{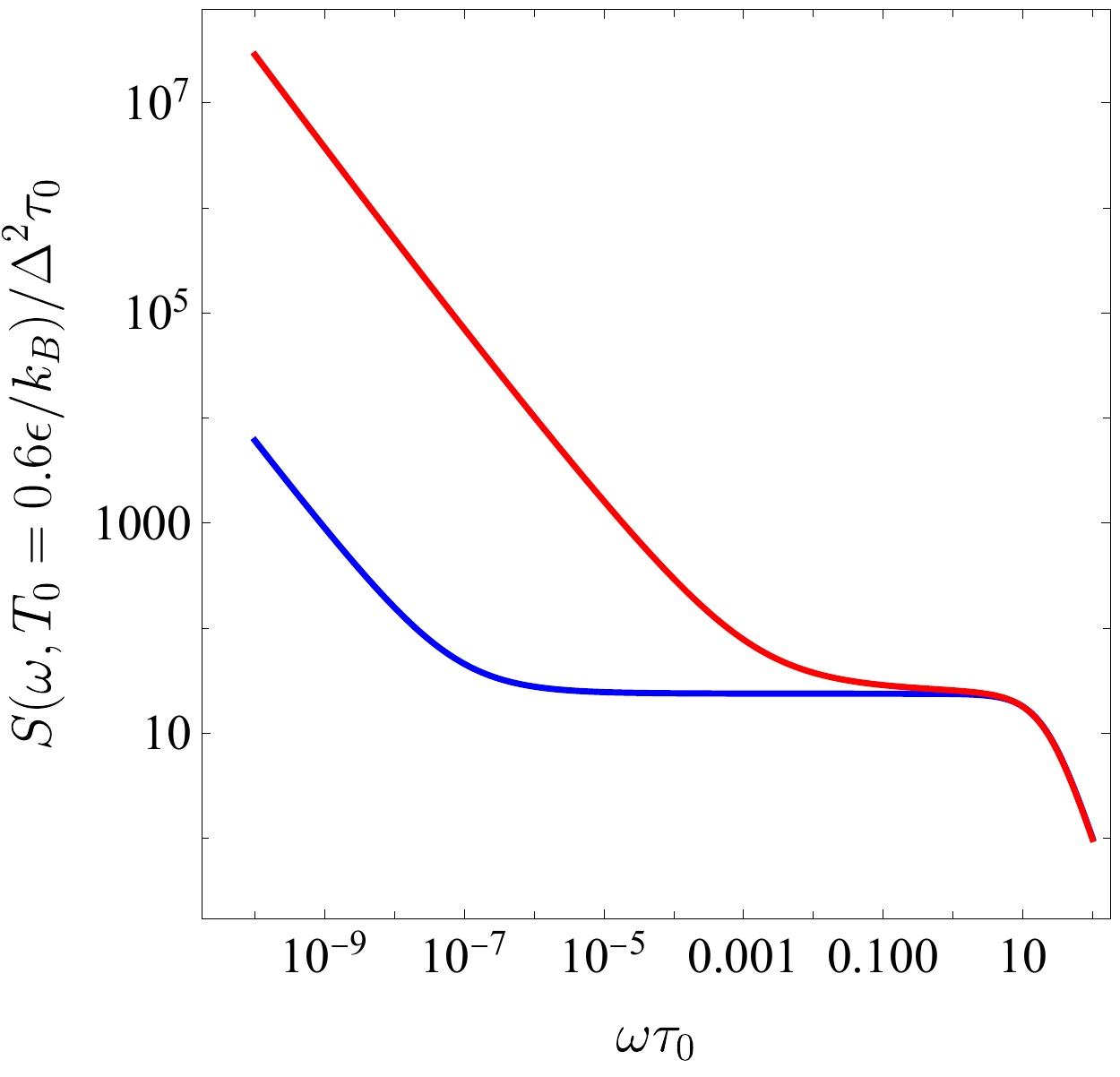}
	\caption{Plots of $S(\omega,T_0)$ as a function of the frequency $\omega$ for constant temperature $k_BT_0/\epsilon=0.2$ (left panel), $0.4$ (middle panel), and $0.6$ (right panel).  The blue curves use the model of Ahn {\it el. al.} \cite{PhysRevB.103.L041304}, and the red curves use our model.}
	\label{fig:S_ConstT}
\end{figure*}

\begin{figure*}
	\centering
	\includegraphics[width=0.32\textwidth]{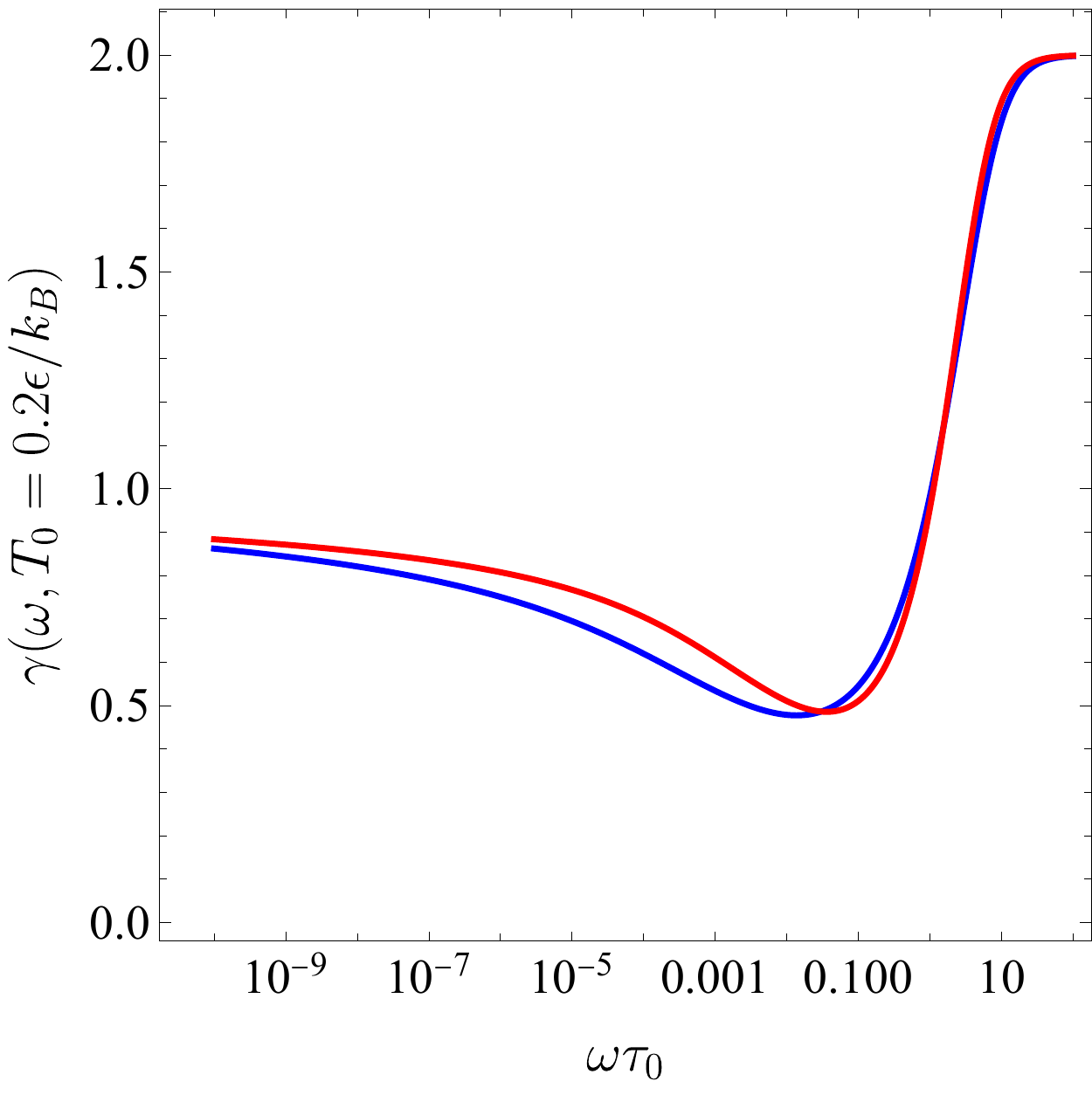}
	\includegraphics[width=0.32\textwidth]{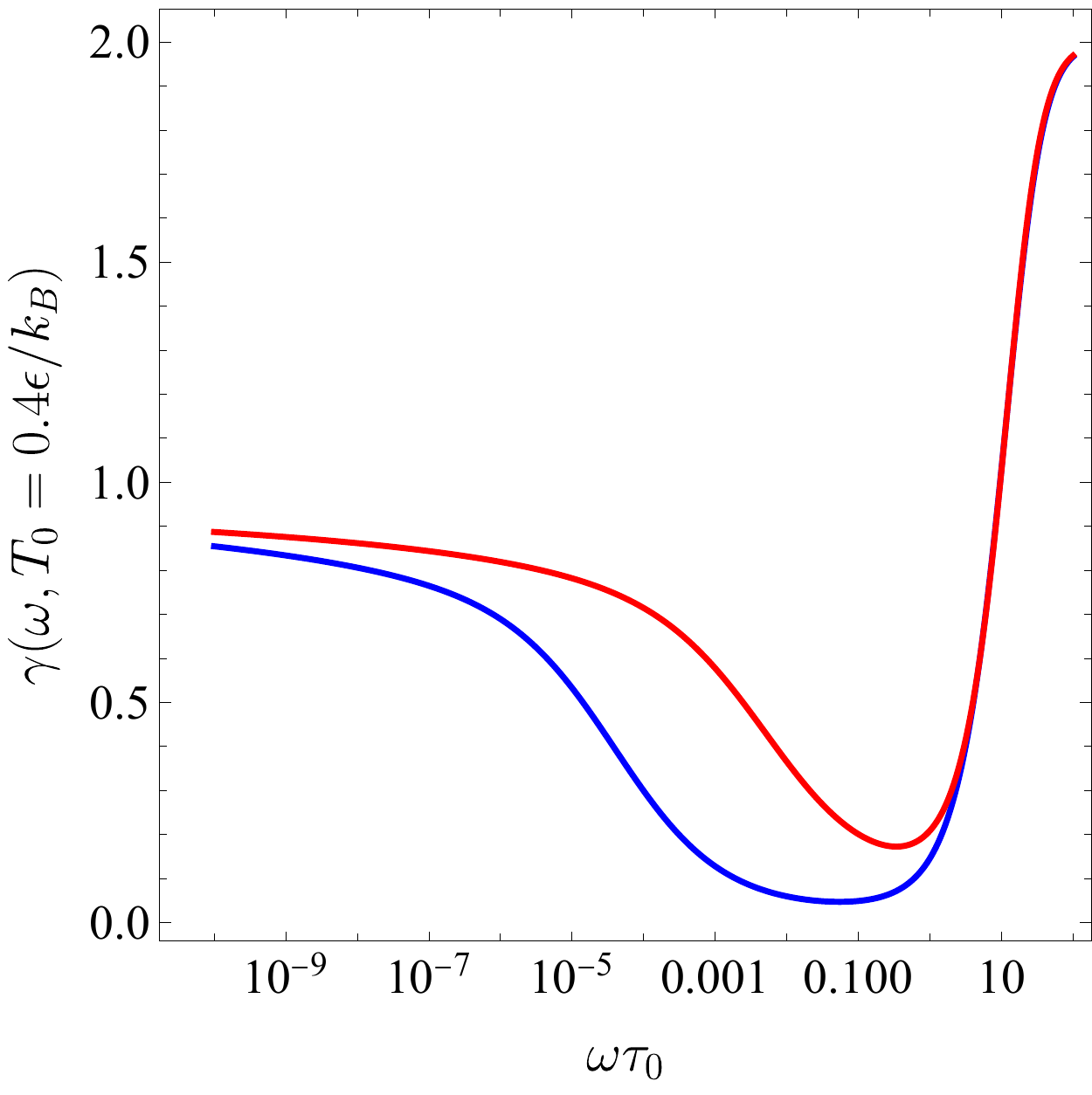}
	\includegraphics[width=0.32\textwidth]{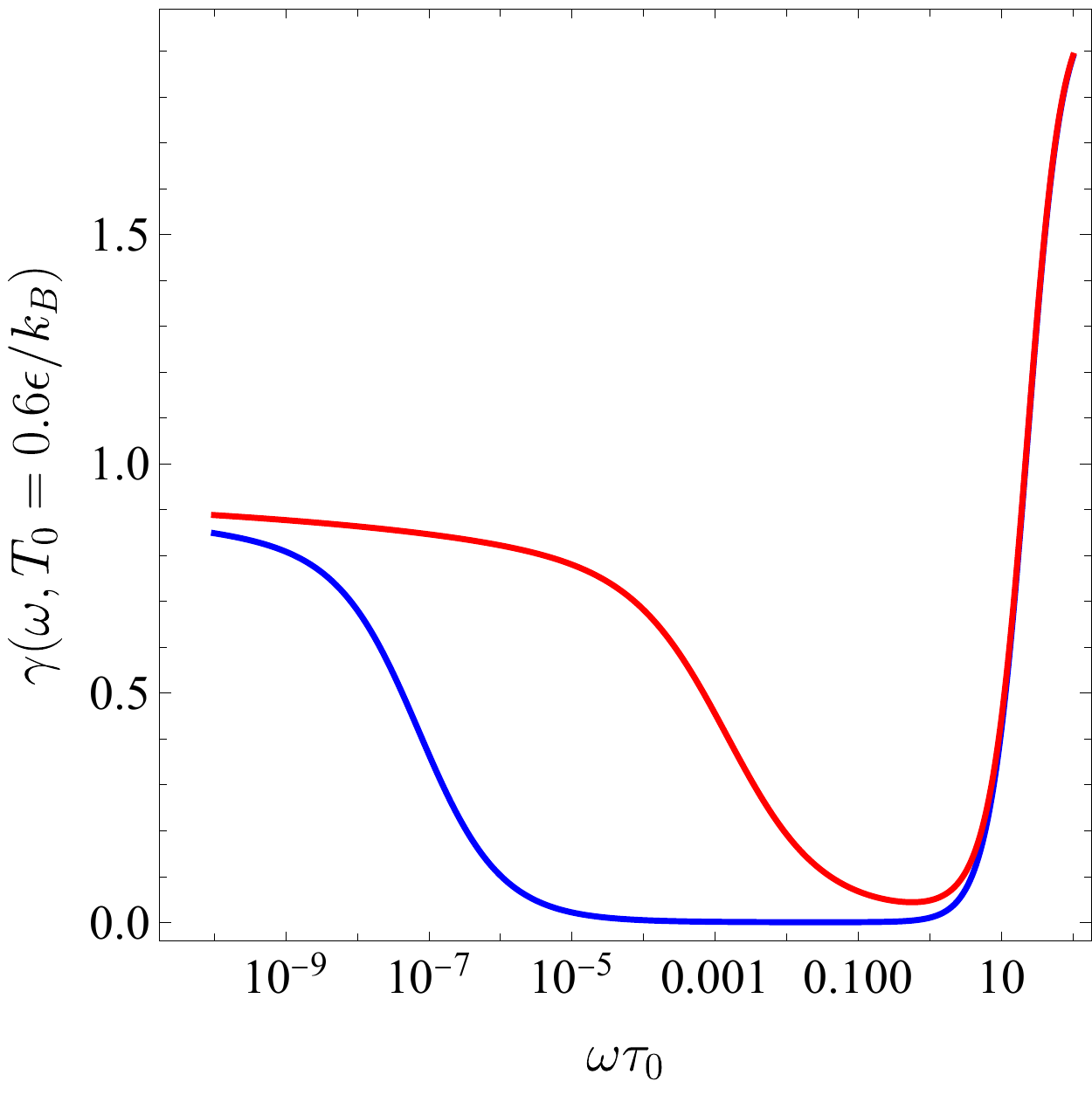}
	\caption{Plots of $\gamma(\omega,T_0)$ as a function of the frequency $\omega$ for constant temperature $k_BT_0/\epsilon=0.2$ (left panel), $0.4$ (middle panel), and $0.6$ (right panel).  The blue curves use the model of Ahn {\it el. al.} \cite{PhysRevB.103.L041304}, and the red curves use our model.}
	\label{fig:gamma_ConstT}
\end{figure*}

\begin{figure*}
	\centering
	\includegraphics[width=0.32\textwidth]{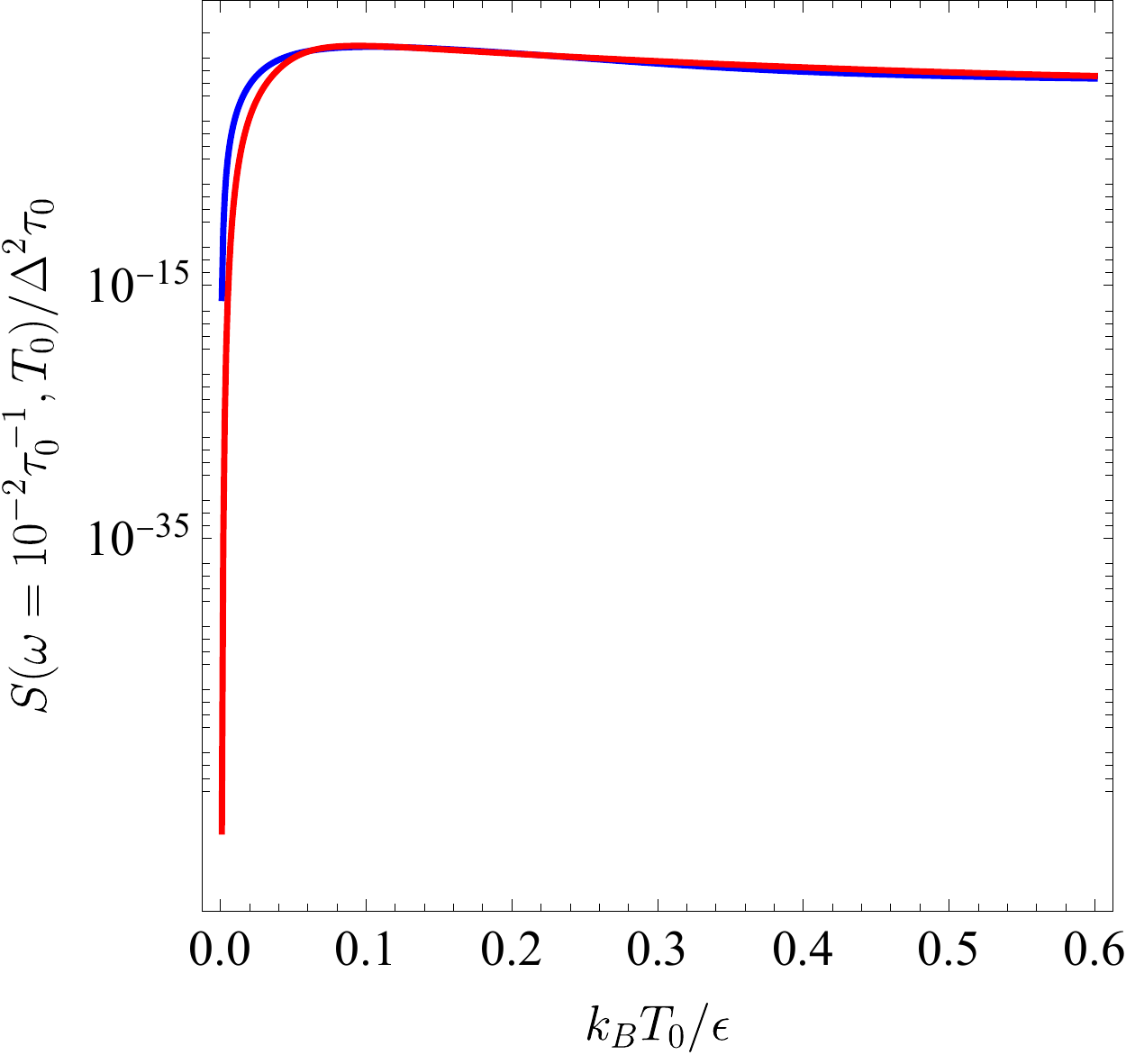}
	\includegraphics[width=0.32\textwidth]{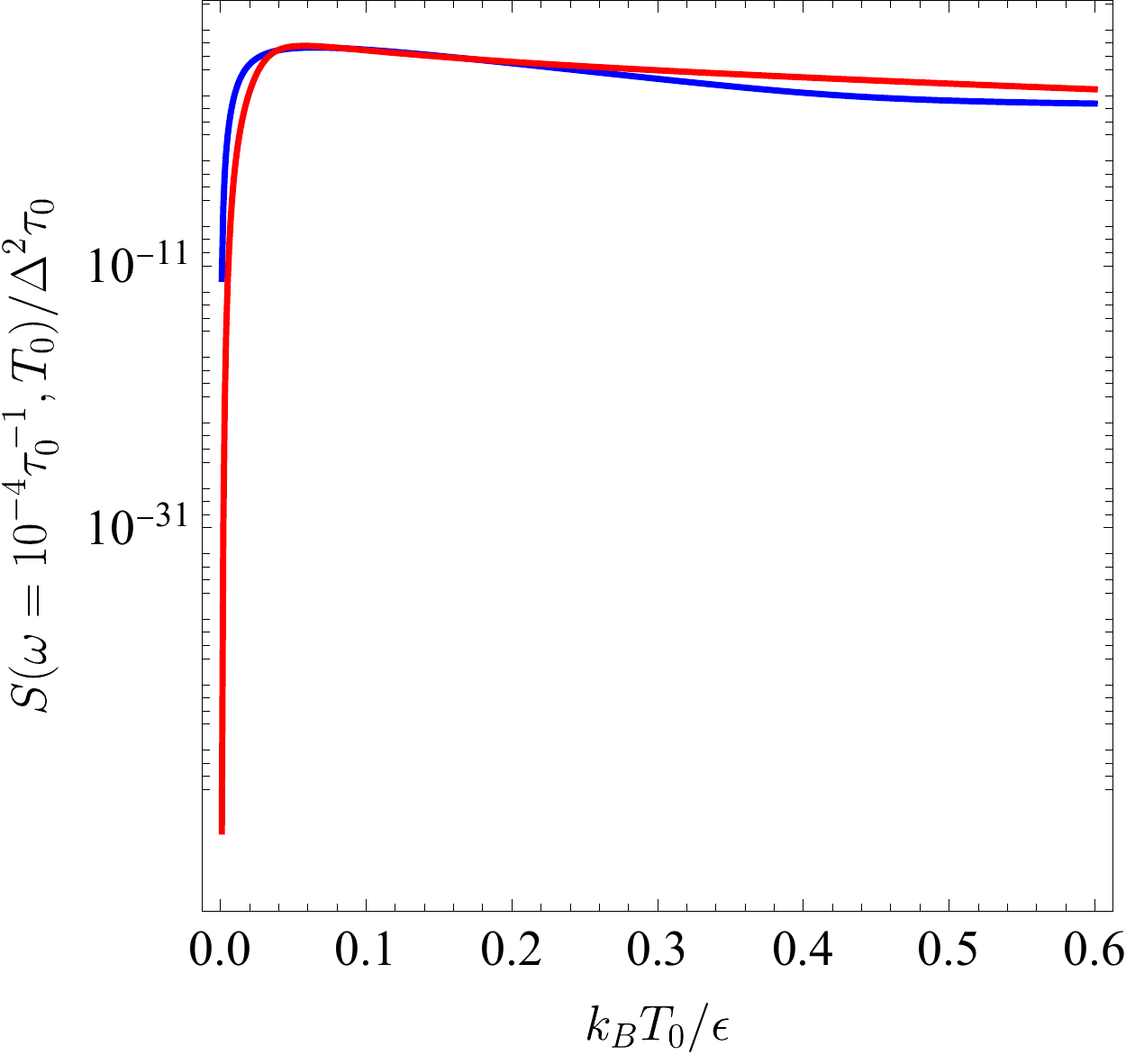}
	\includegraphics[width=0.32\textwidth]{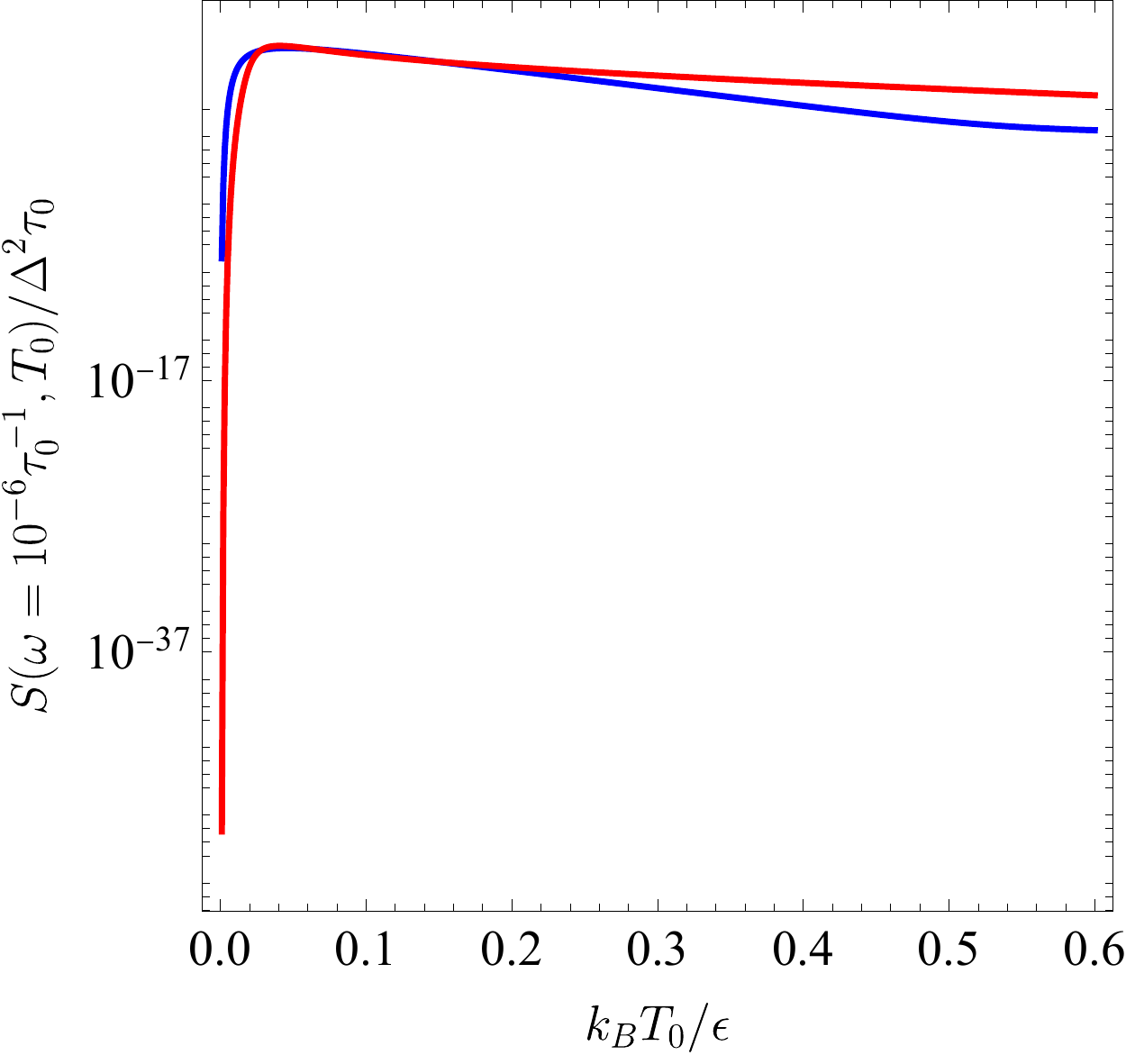}
	\caption{Plots of $S(\omega,T_0)$ as a function of the full electron bath temperature $T_0$ for constant frequency $\omega\tau_0=10^{-2}$ (left panel), $10^{-4}$ (middle panel), and $10^{-6}$ (right panel).  The blue curves use the model of Ahn {\it el. al.} \cite{PhysRevB.103.L041304}, and the red curves use our model.}
	\label{fig:S_ConstOmega}
\end{figure*}

\begin{figure*}
	\centering
	\includegraphics[width=0.32\textwidth]{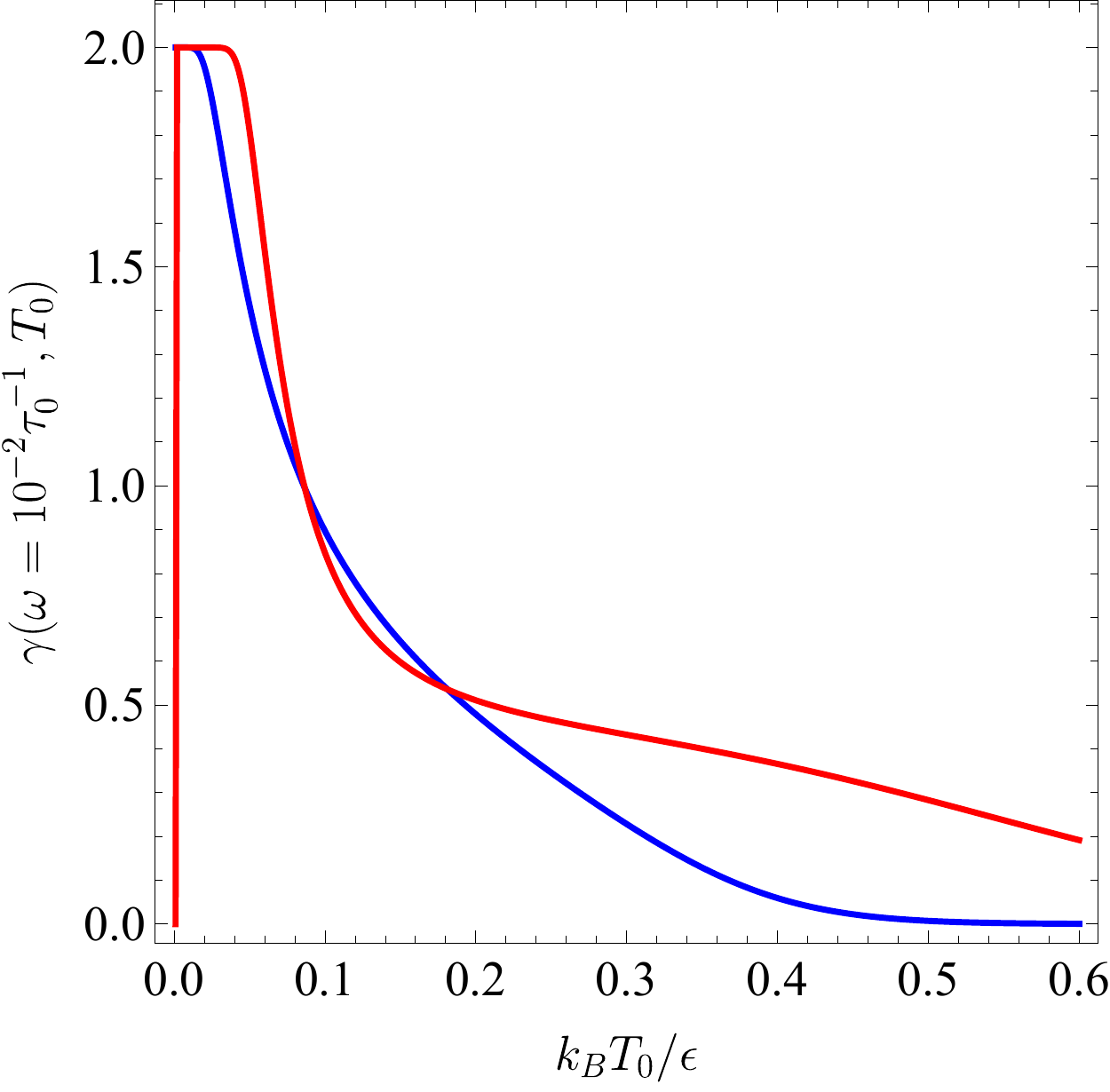}
	\includegraphics[width=0.32\textwidth]{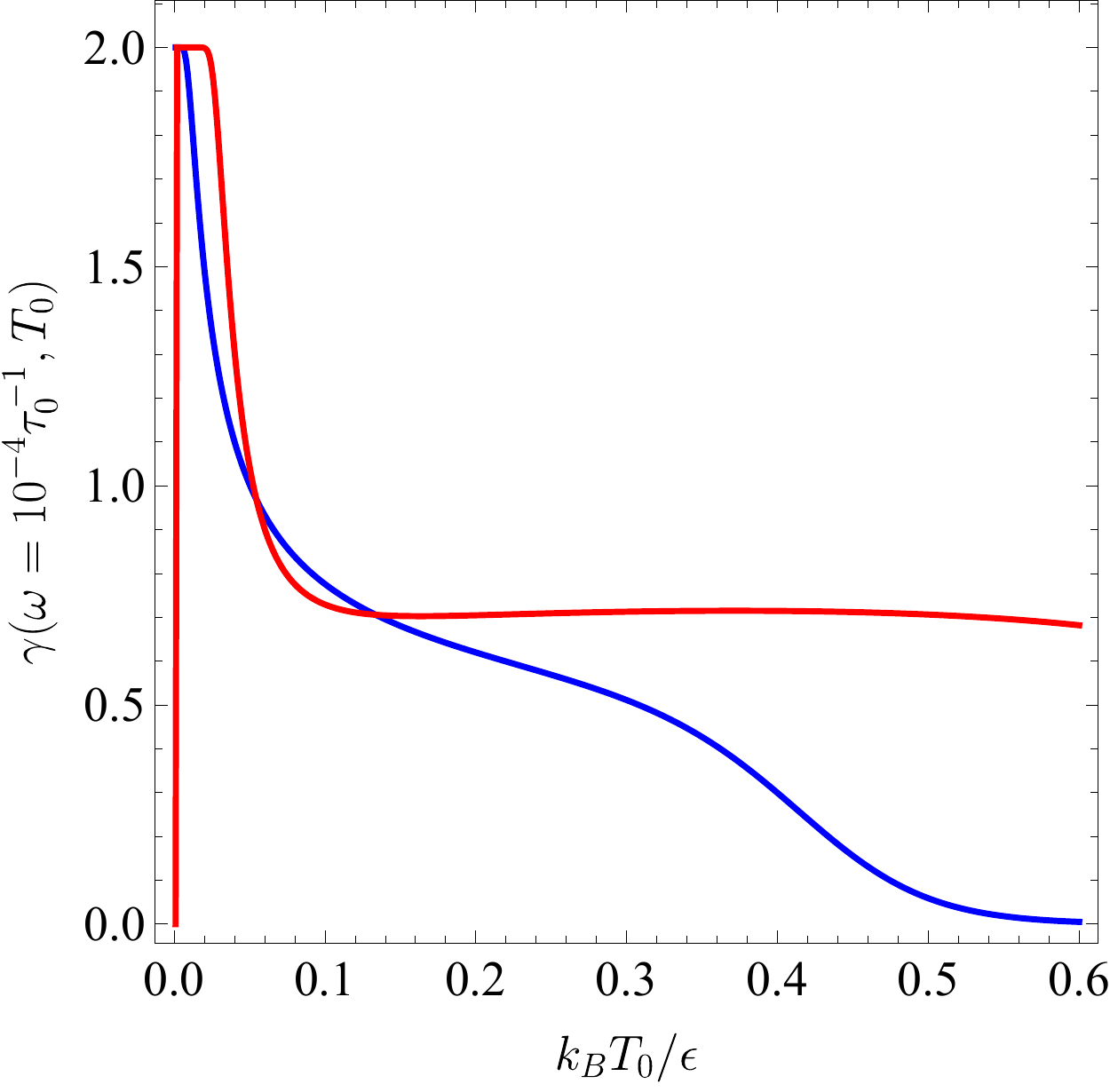}
	\includegraphics[width=0.32\textwidth]{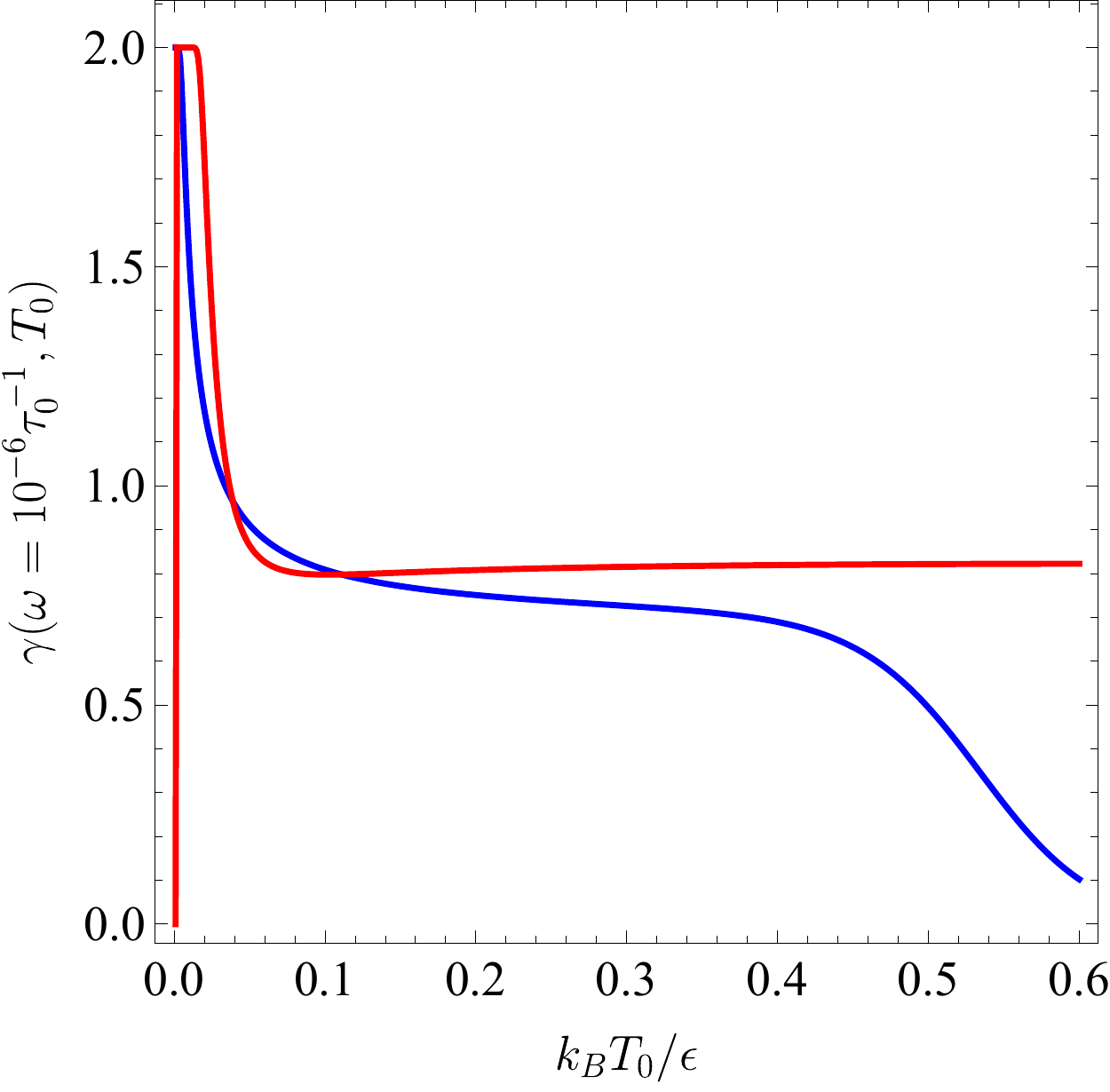}
	\caption{Plots of $\gamma(\omega,T_0)$ as a function of the full electron bath temperature $T_0$ for constant frequency $\omega\tau_0=10^{-2}$ (left panel), $10^{-4}$ (middle panel), and $10^{-6}$ (right panel).  The blue curves use the model of Ahn {\it el. al.} \cite{PhysRevB.103.L041304}, and the red curves use our model.}
	\label{fig:gamma_ConstOmega}
\end{figure*}

\begin{figure*}
	\centering
	\includegraphics[width=0.4\linewidth]{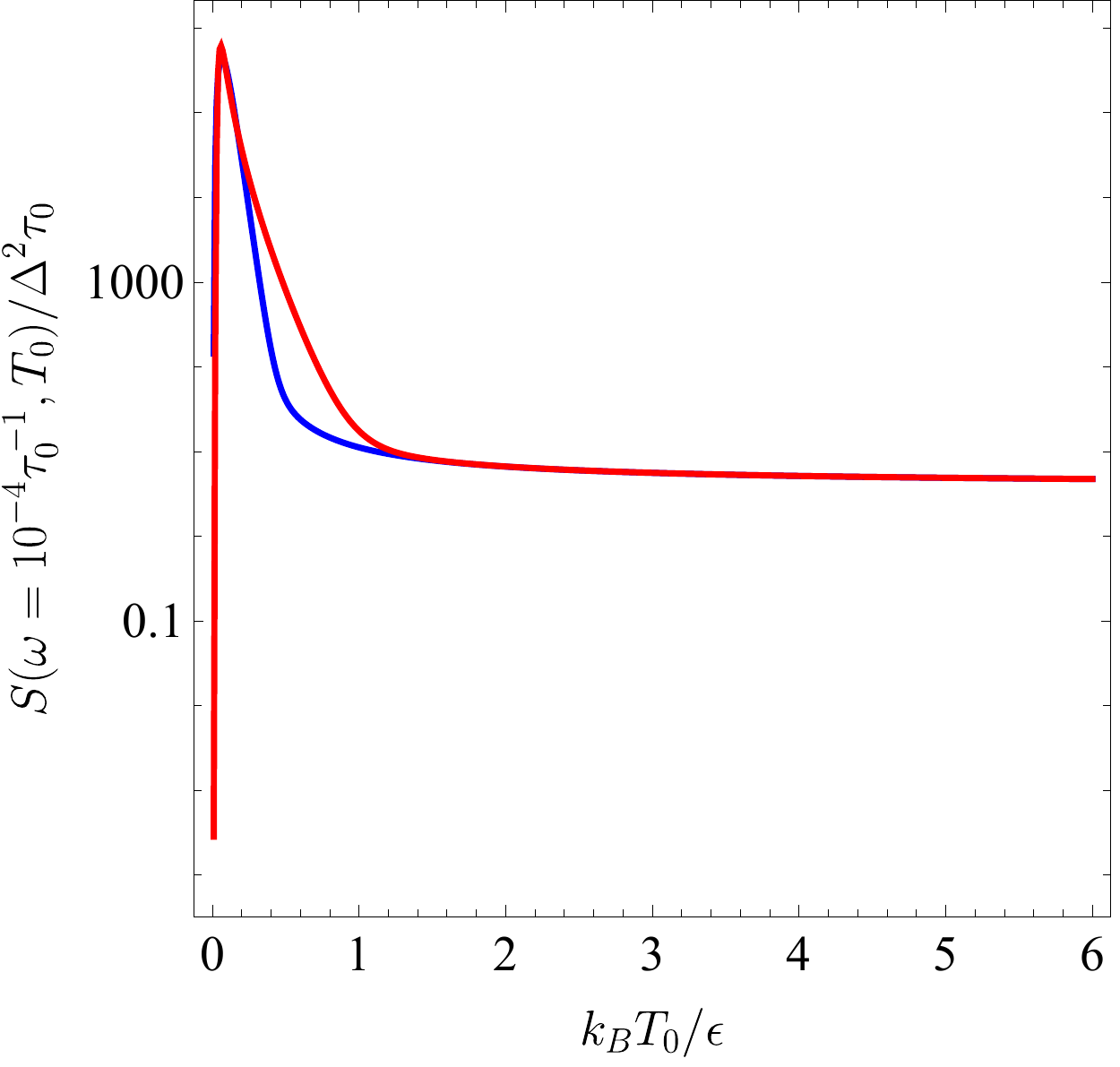}
	\includegraphics[width=0.4\linewidth]{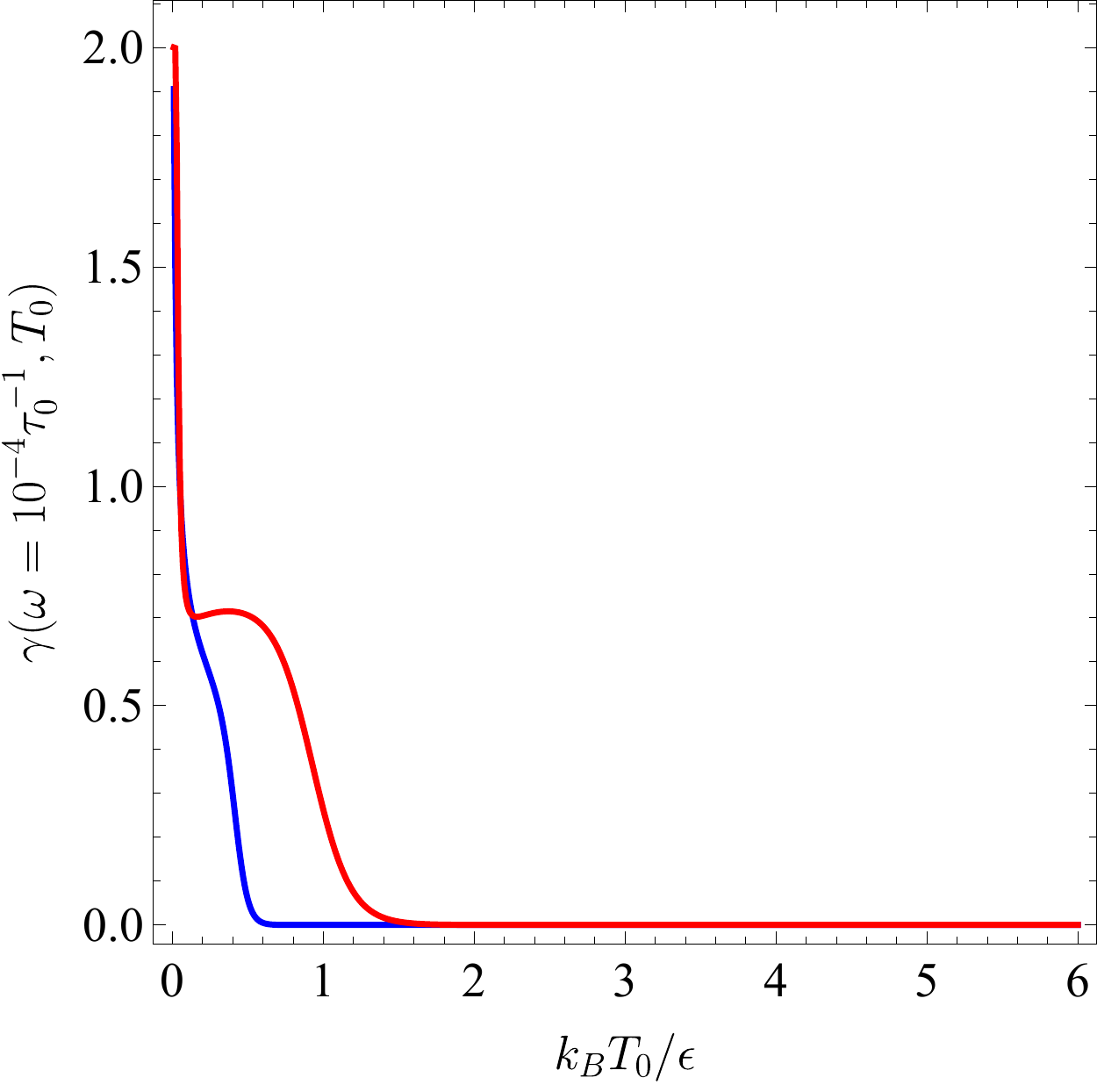}
	\caption{Plots of the noise spectrum $S(\omega,T_0)$ (left panel) and $\gamma(\omega,T_0)$ (right panel) as functions of the full electron bath temperature $T_0$ for constant frequency $\omega\tau_0=10^{-4}$ for high temperatures (up to $k_BT_0/\epsilon=6$).}
	\label{fig:ConstOmegam4_HT}
\end{figure*}

We plot $S$ and $\gamma$, respectively, for fixed temperature as functions of frequency in Figs.~\ref{fig:S_ConstT} and \ref{fig:gamma_ConstT} and vice versa in Figs.~\ref{fig:S_ConstOmega} and \ref{fig:gamma_ConstOmega}.  One feature that we notice is that, especially for $k_BT_0/\epsilon=0.4$ or $0.6$, our model yields values of $S(\omega,T_0)$ that are several orders of magnitude larger than those yielded by Ahn {\it el. al.}'s model \cite{PhysRevB.103.L041304}.  This result appears at first to contradict the fact that our model would actually agree with Ahn {\it el. al.}'s at higher temperatures.  However, the temperatures shown in those plots are still in the low-temperature regime within which we expect the two to differ; if we extend them to $k_BT_0/\epsilon=6$, as shown in Fig.~\ref{fig:ConstOmegam4_HT}, then we find that the two models agree at high temperatures ($k_BT_0/\epsilon\geq1.5$), as expected.  We also note that the overall qualitative behavior of the two models, as described by $\gamma$, is the same, with $1/\omega^{0.9}$ behavior at very low frequencies, which is close to the Dutta {\it el. al.} result of $1/\omega$ \cite{PhysRevLett.43.646} and $1/\omega^2$ at high frequencies, with the crossover between these two regimes set by the temperature.  However, the frequencies at which we see crossovers from one behavior to another will depend on the model.  We also see $1/\omega^2$ behavior at low temperature in both models.  The high-frequency limit is consistent with experimental observations, e.g., Ref.~\cite{PhysRevB.100.165305}.


\begin{figure*}[tbh]
	\centering
	\includegraphics[width=0.4\linewidth]{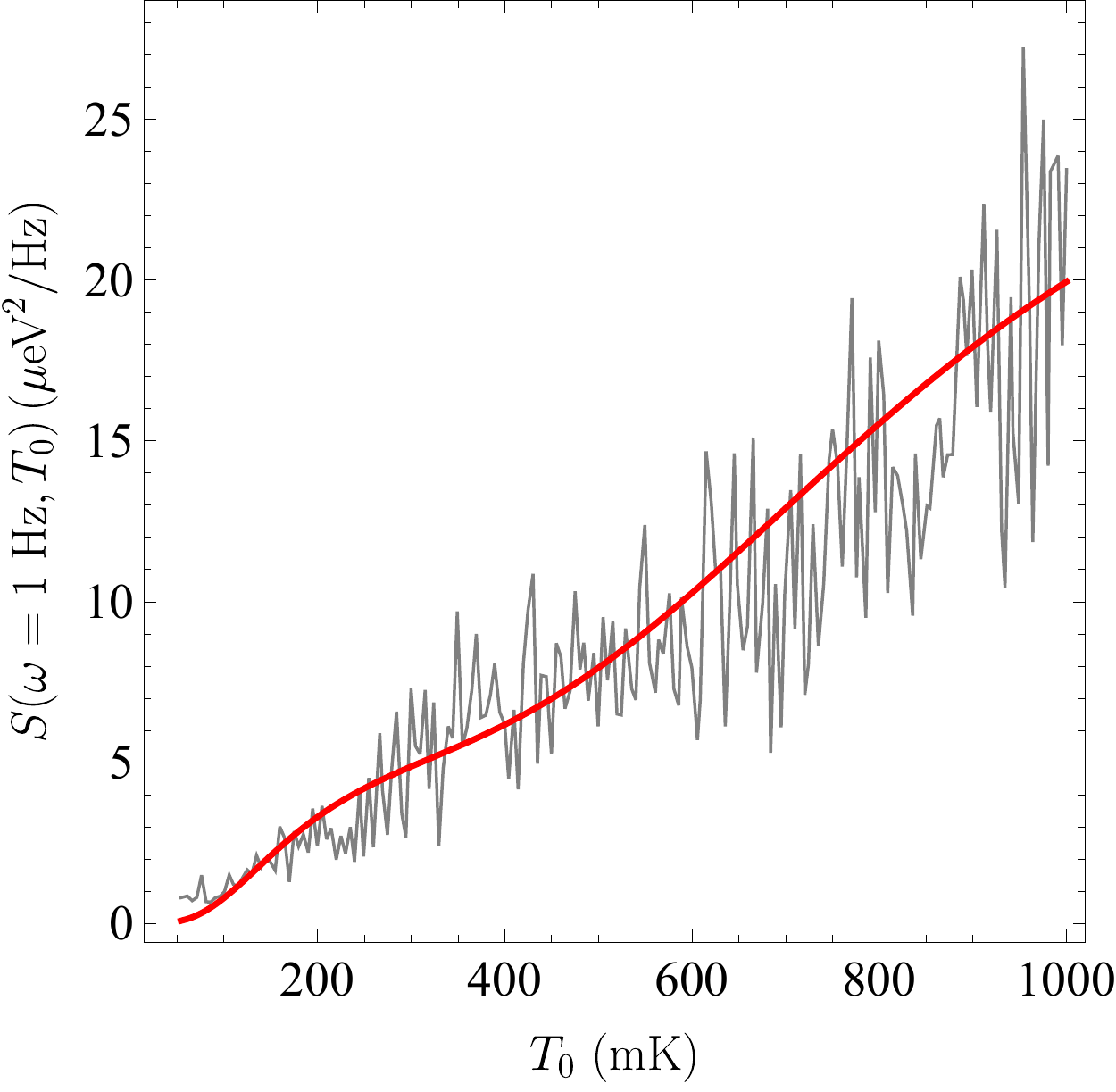}
	\includegraphics[width=0.4\linewidth]{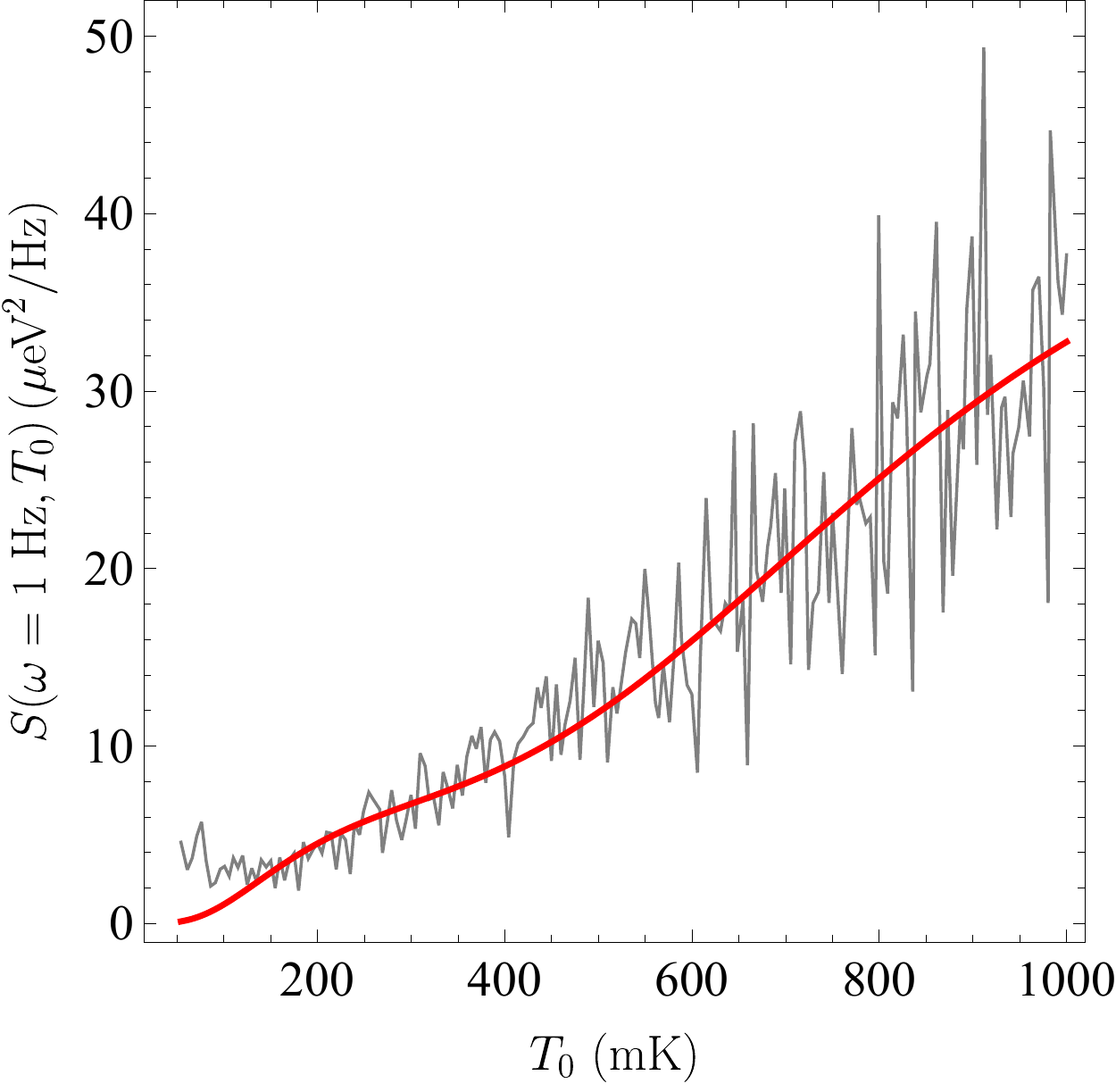}
	\caption{Fit of the noise spectrum measured in Ref.~\cite{PhysRevB.100.165305} as a function of the full electron bath temperature $T_0$ for constant frequency $\omega=1\text{ Hz}$ to two TLFs using the exact temperature distribution, Eq.~\eqref{Eq:TempDistExact}.  The data comes from Figs.~2(u) (left) and 2(v) (right) in the Supplemental Material for Ref.~\cite{PhysRevB.100.165305}, and the fit parameters are found in Table \ref{tab:Fit_JNData_Params}.}
	\label{fig:Fit_JNData}
\end{figure*}

We now perform a fit of the experimentally observed noise spectra of Ref.~\cite{PhysRevB.100.165305}, which considered a device consisting of two quantum dots, to our own formulas.  We assume that the qubits were affected by two TLFs, each with a different strength $\Delta_i$, activation energy $\epsilon_i$, and switching time $\tau_{0,i}$; however, both interact with subbaths with the same area $A=\pi r^2$.  We present a plot of the fit along with the data in Fig.~\ref{fig:Fit_JNData} [Supplemental Material for Ref.~\cite{PhysRevB.100.165305}, Figs.~2(u) (left dot) and 2(v) (right dot)] and the values of the parameters used in this fit in Table \ref{tab:Fit_JNData_Params}.  We see that the fit to the data is excellent over all but the lowest temperatures (below $1.75$ mK).  We also note that the same activation energies and switching times fit both data sets; the only difference is in the strengths.  This suggests that the noise spectra observed in both quantum dots is due to the same two TLFs, as the differences in strength may be explained simply by the positions of the TLFs relative to the quantum dots.
\begin{table}[tbh]
\begin{tabular}{|c|c|c|c|c|}
	\hline
{\bf Defect}	& \multicolumn{2}{c|}{$1$} & \multicolumn{2}{c|}{$2$} \\
	\hline
{\bf Dot}	& Left & Right & Left & Right \\
	\hline
$\Delta$ ($\mu\text{eV}$) & $0.064088$ & $0.588603$ & $0.0746267$ & $0.770345$ \\
	\hline
$\epsilon$ (meV) & \multicolumn{2}{c|}{$0.15778$} & \multicolumn{2}{c|}{$0.6528$} \\
	\hline
$\tau$ (ms)	& \multicolumn{2}{c|}{$110.23$} & \multicolumn{2}{c|}{$57.104$} \\
	\hline
$r$ (nm)	& \multicolumn{4}{c|}{$18.28$} \\
	\hline
\end{tabular}
\caption{Table of parameters used to produce the fits in Fig.~\ref{fig:Fit_JNData}.}
\label{tab:Fit_JNData_Params}
\end{table}

\section{Conclusion} \label{Sec:Conclusion}
We analyzed the effects of dropping the assumption of an electronic heat capacity that is constant in temperature implicit in Eq.~\eqref{Eq:SigmaSB_CVConst} on the results derived by Ahn {\it et. al.} in Ref.~\cite{PhysRevB.103.L041304} for the behavior of the noise power spectrum $S(\omega,T_0)$.  We began by deriving the exact temperature distribution under the assumption that the energies have a truncated Gaussian distribution, and then deriving the corresponding approximate Gaussian distribution for the subbath temperature, finding a relationship between the standard deviation of the energy distribution $\sigma_E$ and that of the effective Gaussian temperature distribution $\sigma_{sb}$.  We find that, for low temperatures, which are the most relevant for experiments, $\sigma_{sb}\propto T_0^{3/4}$, rather than $\sqrt{T_0}$ as in Ahn {\it et. al.}'s work.  We then turned our attention to the noise power spectrum $S(\omega,T_0)$.  We first determined its low-temperature behavior analytically, and found that it now goes as $e^{-C/T_0^{3/8}}$, in contrast with $e^{-C'/T_0^{1/3}}$ as found in Ref.~\cite{PhysRevB.103.L041304}.  We confirmed numerically that the Gaussian approximation to the subbath temperature distribution used in deriving these results is valid.  We then performed a numerical calculation of the noise power spectrum both using Ahn {\it et. al.}'s approach and our approach, finding that the results may differ by several orders of magnitude at low temperatures, but that they agree at high temperatures, as expected.

We should note, however, that we see no qualitative differences between our model and that of Ref.~\cite{PhysRevB.103.L041304}.  In both models, we find that, for low frequencies, $S(\omega,T_0)\propto 1/\omega^{0.9}$, while $S(\omega,T_0)\propto 1/\omega^2$ at high frequencies.  The frequency at which we observe this crossover depends on the temperature, with the approximate $1/\omega^{0.9}$ dependence persisting over a larger range of frequencies at higher temperatures.  The only change that we see between the two models is the frequency at which this crossover happens.  These dependences of $S(\omega,T_0)$ on $\omega$ are consistent with experimental observations.

Finally, we performed a fit to the experimental data of Ref.~\cite{PhysRevB.100.165305} taken from a two-quantum-dot device to the spectrum produced by two TLFs using our model.  We find that it is possible to produce an excellent fit of our model to the data.  We saw that the same activation energies and switching times produced such excellent fits---only the strengths of the two TLFs differed for the two dots.  This implies that the noise spectra observed in both dots can be explained by the same two TLFs, as the differences in TLF strength may be explained by the relative positions of the two TLFs to the two dots.  We thus find that, as was concluded in Ref.~\cite{PhysRevB.103.L041304}, the observed noise spectrum may be explained by just two TLFs, and that an ensemble of TLFs is not needed.

\acknowledgments
This work is supported by the Laboratory for Physical Sciences.  We thank Professor John Nichol of the University of Rochester for providing us the experimental data found in Ref.~\cite{PhysRevB.100.165305}, which we used to perform the fits shown in Fig.~\ref{fig:Fit_JNData}.

\bibliography{ChargeNoise_LinSH}

\end{document}